\documentclass[conference]{IEEEtran}

\usepackage[letterpaper, top=0.75in, bottom=1in, left=0.625in, right=0.625in]{geometry}
\usepackage{cite}
\usepackage{amsmath,amssymb,amsfonts}
\usepackage{algorithmic}
\usepackage{graphicx}
\usepackage{subcaption}
\usepackage{textcomp}
\usepackage{xcolor}
\usepackage{multirow}
\usepackage{dirtytalk}
\usepackage{hyperref}

\begin{document}
\title{Interference Prediction Using Gaussian Process Regression and Management Framework for Critical Services in Local 6G Networks
\thanks{The research leading to this paper was supported by the Research Council of Finland (former Academy of Finland) through the projects \href{https://www.6gflagship.com/}{6G Flagship} (grant number: 369116), and 6G-ConCoRSe (grant number: 359850).}}

 \author{\IEEEauthorblockN{Syed Luqman Shah, Nurul~Huda~Mahmood, and Matti~Latva-aho, \textit{Fellow, IEEE}
 }\\
 \IEEEauthorblockA{6G Flagship, Centre for Wireless Communications (CWC), University of Oulu, Oulu FI-90014, Finland}
\{syed.luqman, nurulhuda.mahmood, matti.latva-aho\}@oulu.fi
 }
 \maketitle

\begin{abstract}
Interference prediction and resource allocation are critical challenges in mission-critical applications where stringent latency and reliability constraints must be met. This paper proposes a novel Gaussian process regression (GPR)-based framework for predictive interference management and resource allocation in future 6G networks. Firstly, the received interference power is modeled as a Gaussian process, enabling both the prediction of future interference values and their corresponding estimation of uncertainty bounds. Differently from conventional machine learning methods that extract patterns from a given set of data without any prior belief, a Gaussian process assigns probability distributions to different functions that possibly represent the data set which can be further updates using Bayes' rule as more data points are observed. For instance, unlike deep neural networks, the GPR model requires only a few sample points to update its prior beliefs in real-time. Furthermore, we propose a proactive resource allocation scheme that dynamically adjusts resources according to predicted interference. The performance of the proposed approach is evaluated against two benchmarks prediction schemes, a moving average-based estimator and the ideal genie-aided estimator. The GPR-based method outperforms the moving average-based estimator and achieves near-optimal performance, closely matching the genie-aided benchmark.

\end{abstract}


\section{Introduction}
\label{Sec: Introduction}
In sixth-generation (6G) wireless networks, critical services for applications such as the tactile Internet~\cite{Tactile_Internet}, autonomous vehicles~\cite{vehicular_networks}, and industrial automation~\cite{factory_automation} demand stringent connectivity and reliability requirements. To address these requirements, IMT-2030 identifies the hyper-reliable low-latency communication (HRLLC) use case, which define stringent constraints on latency, specified within the range of $[0.1\text{ms}, 1\text{ms}]$, and reliability, targeted within the interval $[10^{-5}, 10^{-7}]$~\cite{IMT_2030}.

Meeting these stringent requirements is challenging, particularly in environments with high interference. Predictive interference management has been proposed as a key enabler for ensuring the quality of service (QoS) in 6G networks~\cite{6}. By forecasting interference reliably, networks can proactively allocate resources to maintain the required performance for mission-critical applications in an efficient manner~\cite{Samitha_paper, Samitha_Ref, 7_1, Nurul_paper}. 

Interference prediction and resource allocation for critical services are gaining attention, with two main approaches: stochastic methods and machine learning (ML)-based techniques. Stochastic methods, including conventional link adaptation, predict future interference using historical samples, often relying on mean values. Such an approach overlooks large fluctuations resulting in suboptimal predictions~\cite{7}. More sophisticated approaches, such as autoregressive moving-average (MA) models, improve accuracy by utilizing the entire interference distribution~\cite{7_1, 8}. In~\cite{Nurul_paper}, a discrete-time Markovian interference model is developed to forecast interference state transitions, enabling proactive resource allocation for wireless critical applications. In vehicular networks, predictive models have been used to mitigate packet collisions by anticipating interference~\cite{10}.

On the other hand, ML-based approaches particularly those using neural networks, improve prediction accuracy by leveraging large datasets to learn interference patterns. Interference-aware resource allocation for efficient deep learning (DL) on graphics processing units is explored in~\cite{9}, while models like long short-term memory have been applied to wireless interference prediction~\cite{Samitha_Ref}. Further advancements include the use of deep neural networks (DNNs) with transformer architecture, as discussed in~\cite{Samitha_paper}. However, ML-based methods have three key drawbacks: they do not provide confidence levels for predictions, they cannot easily incorporate prior knowledge, and they typically require large training datasets.

This work introduces a GPR-based framework for predicting interference power and enabling proactive resource allocation in local 6G deployments. Local 6G networks, which operate in confined environments such as industrial and healthcare settings, require HRLLC services due to their stringent application demands~\cite{Samitha_paper}. Compared to traditional stochastic models, which rely on oversimplified assumptions and DNN-based methods that require large datasets and lack interpretability, the GPR-based approach offers several key advantages~\cite{gaussian_IEEE_magzine, Gaussian_Book}: (i) its non-parametric nature adapts to complex, non-linear interference dynamics without a predefined model structure; (ii) its data efficiency allows accurate predictions with minimal training data, ideal for data-constrained 6G environments~\cite{Gaussian_Regression_Journal}; (iii) its real-time adaptability ensures dynamic updates as new data is available, responding to evolving interference conditions; (iv) its transparency and interpretability enable hyperparameter tuning and the integration of prior knowledge, overcoming limitations of black-box ML methods; and (v) its uncertainty quantification provides confidence intervals for predictions, facilitating informed resource allocation and meeting the rigorous service quality requirements of HRLLC applications~\cite{gaussian_IEEE_magzine, Gaussian_Regression_Journal}.

The primary contributions of this work are threefold. First, we develop and implement a GPR-based model for predicting received interference power, specifically tailored for local 6G networks. Second, we integrate the predicted interference power into a proactive resource allocation framework, enabling efficient resource distribution to meet the stringent service requirements of mission-critical applications. Third, we conduct a performance evaluation of the proposed GPR-based model, comparing it against two benchmark methods MA-based estimation and genie-aided estimation, with results presented in terms of achieved outage versus target outage\footnote{To facilitate reproducibility, the simulation codes are publicly accessible at: \url{https://github.com/Syed-Luqman-Shah-19/GPR-IEEEWCNC2025}.}.

Following the introduction, review of existing techniques, and outline of the contributions in Section~\ref{Sec: Introduction}, the system model is detailed in Section~\ref{Sec: System Model}. Section~\ref{sec:gp_interference_prediction} presents the non-parametric GPR model for interference power prediction. A proactive resource allocation strategy is introduced in Section~\ref{sec:resourceAllocation}. Section~\ref{Sec: Results and Discussion} presents the results, while Section~\ref{Sec: Conclusions and Future} discusses potential future directions and concludes the paper.

\begin{figure}[t]
    \centering
    \includegraphics[width=0.8\linewidth]{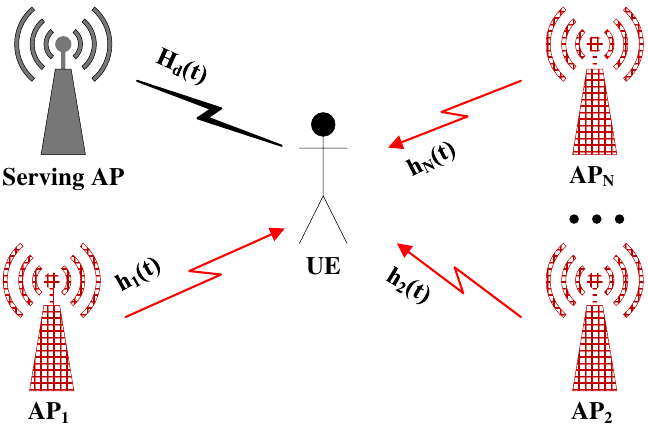}
    \caption{Proposed indoor wireless system model depicting a desired signal and $N$ interfering signals to a UE.}
    \label{fig:system_model}
\end{figure}

\section{System Model} 
\label{Sec: System Model}

We consider the downlink communication of a local 6G network deployed in an indoor mission-critical environment, such as a factory or industrial setting. The system comprises a serving access point (AP) communicating with user equipment (UE), surrounded by $N$ interfering APs, all operating within the same frequency band, as illustrated in Fig.~\ref{fig:system_model}. Due to the stationary nature of the devices in this setting, the system is assumed to be quasi-static during each transmission period. The communication channels are subject to multipath propagation, resulting in Rayleigh fading for all links due to reflections from surrounding structures. The $N$ interfering APs introduce co-channel interference, which may degrade the signal quality received by the UE.

The link between the serving AP and the UE is characterized by an average signal-to-noise ratio (SNR), ($\bar{\gamma}$). Each interfering AP contributes an interference-to-noise ratio (INR), ($\gamma_i$), for $i = 1, 2, \dots, N$, with maximum and minimum values $\gamma_{\text{max}}$ and $\gamma_{\text{min}}$, respectively. We assume that the UE connects to the AP with the strongest signal, where $\bar{\gamma} > \gamma_{\text{max}}$, and that there is no cooperation among the interfering APs (e.g., power control or coordinated beamforming).

The received signal at the UE, $R_r(t)$, is expressed as,

\begin{equation}
    R_r(t) = \sqrt{P_d} h_d(t) s_t(t) + \sum_{i=1}^{N} \sqrt{P_i} h_i(t) s_i(t) + n(t),
\end{equation}

where $P_d$ is the transmitted power from the serving AP, $h_d(t)$ is the channel coefficient of the desired link, $s_t(t)$ is the transmitted symbol, $P_i$ is the transmitted power from the $i$-th interferer, $h_i(t)$ is the channel coefficient of the $i$-th interfering link, and $n(t) \sim \mathcal{CN}(0, N_0)$ represents AWGN with noise power spectral density $N_0$. The total interference power $I(t)$ received by the UE at time $t$ is given as,
\begin{equation}
    I(t) = \sum_{i=1}^{N} P_i |h_i(t)|^2,
\end{equation}
where $|h_i(t)|^2$ follows an exponential distribution due to Rayleigh fading. Finally, the signal-to-interference-plus-noise ratio (SINR) at the UE, $\delta(t)$, is defined as follows,

\begin{equation}
    \delta(t) = \frac{P_d |h_d(t)|^2}{I(t) + N_0},
\end{equation}
where $P_d |h_d(t)|^2$ represents the desired signal power, and $I(t) + N_0$ denotes the total interference plus noise power.

\section{Interference Power Prediction Using Gaussian Processes}
\label{sec:gp_interference_prediction}

Accurate interference prediction is essential for maintaining reliable communication in mission-critical 6G networks. To achieve this, we model interference power as a stochastic process using GPR, which provides both precise predictions and a robust quantification of prediction uncertainty. The Gaussian Process (GP) framework is inherently non-parametric, making it highly effective for capturing the complex, non-linear dynamics of interference. Different types of covariance functions, or combinations thereof (e.g., radial basis function (RBF), Matérn, and periodic kernels), are employed to encode prior assumptions about the data, allowing the model to adapt to different structural patterns in the interference. These kernels define the GP’s prior distribution and guide the posterior update as new observations become available, enabling dynamic and context-aware predictions.

\subsection{GPs: A Primer}
A GP is a collection of random variables, where any finite subset follows a joint Gaussian distribution. We model the interference power as a GP indexed by time, i.e., $I_t(x) \sim \mathcal{GP}(\mu_{I_t}(x), K(x_i, x_j))$. Here, $\mu_{I_t}(x)$ is the mean function, representing the interference power, while the covariance function $K(x_i, x_j)$ captures the correlation between all the possible pair input points. Two key properties of GPs are central to GPR, marginalization and conditionality~\cite{Gaussian_Book}. Marginalization ensures that any subset of inputs has a multivariate Gaussian distribution, while conditionality enables updates to predictions as new data is observed. For a set of input points $X = [x_1, \dots, x_n] \in \mathbb{R}^d $, the corresponding interference power values $I_t(X) = [I_t(x_1), \dots, I_t(x_n)]$ follow the multivariate Gaussian distribution,

\begin{equation}
    \mathbf{I}_t(X) \sim \mathcal{N}\left(\mu_{I_t}(X), K(X, X)\right),
\end{equation}
where $\mu_{I_t}(X)$ is the mean vector and $K(X, X)$ is the covariance matrix.

\subsection{GPR for Interference Prediction}

GPR models interference power as a stochastic process, where the prior is specified by a zero-mean GP, $\mu_{I_t}(x) = 0$, and a covariance function or kernel $K(x_i, x_j)$, which governs the structure of the function. The hyperparameters of the kernel, such as length-scale $\ell$ and variance $\sigma_f^2$, define the GP’s smoothness, stationarity, isotropy, and variability. Hyperparameter tuning is typically done by maximizing the log marginal likelihood of the observed data~\cite{gaussian_IEEE_magzine, Gaussian_Book}. In this work, the RBF kernel is employed to model correlations within the interference data due to its suitability for capturing smooth, continuous variations. The kernel assumes higher correlation between points closer in the input space, aligning with the spatial and temporal characteristics of interference power in 6G networks. Its flexibility and robustness make it ideal for modeling complex, non-linear interference dynamics while ensuring smooth predictions.

\subsubsection{The RBF Kernel}
The RBF kernel is stationary, relying only on the distance between input points rather than their absolute positions, making it effective for interference prediction in quasi-static environments. The RBF kernel is defined as,
\begin{equation}
    K(x_i, x_j) = \sigma_f^2 \exp \left( -\frac{\|x_i - x_j\|^2}{2\ell^2} \right),
\end{equation}
it models smooth, stationary processes where $\ell$ controls prediction smoothness and $\sigma_f^2$ sets the output scale.

\subsubsection{Posterior Update in GPR}
\label{subsec: PosteriorUpdate}
Given training data $X$ and observed interference power $I_t(X)$, GPR predicts future interference power $I_p(X^*)$ at new points $X^*$. The joint distribution of the observed and unobserved data follows,
\begin{equation}
\label{eq: 6}
    \resizebox{0.85\columnwidth}{!}{$
    \begin{pmatrix} I_t(X) \\ I_p(X^*) \end{pmatrix} \sim \mathcal{N} \left( \begin{pmatrix} \mu_{I_t}(X) \\ \mu_{I_p}(X^*) \end{pmatrix}, \begin{pmatrix} K(X, X) & K(X, X^*) \\ K(X^*, X) & K(X^*, X^*) \end{pmatrix} \right),$}
\end{equation}

where $K(X, X^*)$ is the cross-covariance matrix. Using the properties of the joint Gaussian, the posterior distribution for the predicted interference is Gaussian, with mean and covariance, i.e., $I_p(X^*) | I_t(X) \sim \mathcal{N}\left( \mu_{I_p|I_t}(X^*), \Sigma_{I_p|I_t}(X^*) \right)$,

\begin{align}
\label{Eqn: Posterior mean and variance}
    \resizebox{0.9\columnwidth}{!}{$
    \begin{aligned}
         \textnormal{here }\mu_{I_p|I_t}(X^*) &= \mu_{I_p}(X^*) + K(X^*, X) K(X, X)^{-1} (I_t(X) - \mu_{I_t}(X)), \\
        \textnormal{and }\Sigma_{I_p|I_t}(X^*) &= K(X^*, X^*) - K(X^*, X) K(X, X)^{-1} K(X, X^*).
    \end{aligned}$}
\end{align}

The posterior mean provides the predicted interference values, while the posterior covariance quantifies the uncertainty of the prediction.

\subsubsection{Regression Setting for Interference Prediction}
\label{Sec: Regression Setting using GPR}
In practice, the GP’s conditional distribution is used for predicting future interference values based on past observations. The predictive distribution is given by,
\begin{equation}
\resizebox{0.9\columnwidth}{!}{$
    p(I_p(X^*) | I_t(X), X, X^*) = \mathcal{N}(\mu_{I_p|I_t}(X^*), \Sigma_{I_p|I_t}(X^*)),$}
\end{equation}
where $\mu_{I_p|I_t}(X^*)$ and $\Sigma_{I_p|I_t}(X^*)$ are the posterior mean and covariance as defined in (\ref{Eqn: Posterior mean and variance}), respectively. These predictions are updated as new data becomes available, enhancing the model’s accuracy over time. The predicted interference power for future transmission intervals is given by,
\begin{equation}
    I_p = \mu_{I_p|I_t}(X^*) + \epsilon,
\end{equation}
where $\epsilon$ is Gaussian noise. The 95\% confidence interval for predicted values is,
\begin{equation}
\centering
    I_p \in \left[ \mu_{I_p|I_t} \pm 1.96 \times \sqrt{\text{diag}(\Sigma_{I_p|I_t})}\right],
\end{equation}
where $\text{diag}(\Sigma_{I_p|I_t})$ represents the principal diagonal elements of the variance matrix of the predictions. Continuous updates to the GP model based on new observations improve both the predictions and the confidence intervals, enabling more informed resource allocation in 6G networks.

\section{Resource Allocation}
\label{sec:resourceAllocation}

After predicting the interference power \(I_p\) using GPR, the goal is to proactively manage resource allocation in mission-critical 6G environments. This is achieved by dynamically adjusting transmission resources based on the predicted SINR \(\delta_p\), ensuring QoS requirements are met under varying interference conditions. The predicted SINR, \(\delta_p\), is calculated from the channel state information (CSI) using the transmitted power \(P_d\), predicted interference \(I_p\), and noise power \(N_0\),

\begin{equation}
    \label{eqn:predicted_SINR}
    \delta_p = \frac{P_d |h_d(t)|^2}{I_p + N_0}.
\end{equation}

This predicted SINR \(\delta_p\) directly affects the number of transmittable bits (\(D\)) over the allocated channel uses (\(R\)), subject to a target decoding error probability (\(\varsigma\)) in an AWGN channel. Finite blocklength theory governs this relationship, defining \(D\) as~\cite{Nurul_paper, Samitha_Ref},
\begin{equation}
\label{eqn:Number_of_bits}
D = RC(\delta_p) - Q^{-1}(\varsigma) \sqrt{R V(\delta_p)} + O(\log_2 R),
\end{equation}
where \(C(\delta_p) = \log_2(1 + \delta_p)\) represents the channel capacity, \(V(\delta_p)\) denotes the channel dispersion, and \(Q^{-1}(\varsigma)\) is the inverse Q-function, providing the margin needed to achieve the desired error probability. The term \(O(\log_2 R)\) accounts for higher-order corrections in the blocklength regime.

The number of channel uses \(R\) needed to transmit \(D\) bits with target error probability \(\varsigma\) is estimated by,

\begin{equation}
\label{eqn:channel_usage}
R \approx \frac{D}{C(\delta_p)} + \frac{Q^{-1}(\varsigma)^2 V(\delta_p)}{2 C(\delta_p)^2} \left[ 1 + \sqrt{1 + \frac{4D C(\delta_p)}{Q^{-1}(\varsigma)^2 V(\delta_p)}} \right].
\end{equation}

\begin{figure}[t]
    \centering
    \includegraphics[width=1\linewidth,height=1\textheight,keepaspectratio]{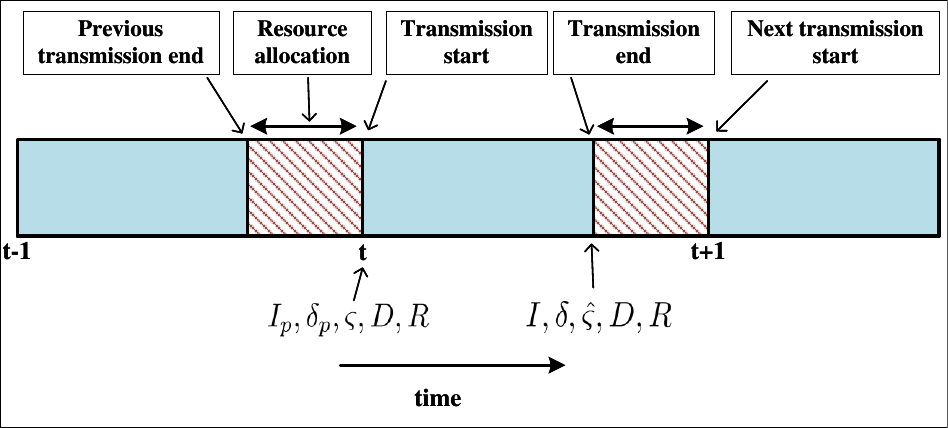}
    \caption{Timeline for the proactive resource allocation scheme.}
    \label{fig:timeline}
\end{figure}

This equation provides an estimate of resource utilization based on the predicted SINR \(\delta_p\) and error target \(\varsigma\). Resource allocation is performed just before transmission, where the CSI from time \( t-1 \) estimates the CSI for time \( t \), and the actual CSI from time \( t \) updates the estimate for \( t+1 \), as depicted in Fig.~\ref{fig:timeline}. Once the transmission takes place, the actual interference power \(I\) is measured at the receiver, allowing the actual SINR \(\delta\) to be computed as,
\begin{equation}
    \label{eqn:actual_SINR}
    \delta = \frac{P_d |h_d(t)|^2}{I + N_0}.
\end{equation}

Typically, \(\delta\) differs from \(\delta_p\) due to unpredictable interference dynamics. The actual decoding error probability \(\hat{\varsigma}\) is then estimated using the finite blocklength formula with the actual SINR \(\delta\),
\begin{equation}
    \label{eqn:decoding_error_actual}
    \hat{\varsigma} \approx Q\left( \frac{R C(\delta) - D}{\sqrt{R V(\delta)}} \right).
\end{equation}

\subsection{Benchmark Resource Allocation Techniques}
To assess the performance of the proposed GPR-based resource allocation by comparing it with two benchmarks, i.e., MA-based and genie-aided estimation.

\subsubsection{Genie-aided Estimation}
The genie-aided estimator represents an optimal, yet impractical, benchmark where perfect knowledge of interference conditions is assumed prior to transmission~\cite{Samitha_paper, Samitha_Ref}. In this idealized scenario, the transmitter has complete foresight of the interference power, allowing perfect interference predictions. Although unrealistic for real-world scenarios, this method provides an upper bound on performance, offering insight into the best possible outcomes of interference prediction techniques~\cite{Nurul_paper}.

\subsubsection{MA-based Estimation}
The MA-based estimator, commonly used in enhanced mobile broadband (eMBB) services, models interference using a first-order infinite impulse response (IIR) filter. The predicted interference power \(I_p\) at the next time step is computed as a weighted sum of the previous interference estimate \(I_{t-1}\) and the current observed interference \(I_t\),
\begin{equation}
    I_p = \alpha I_{t-1} + (1 - \alpha) I_t, \quad 0 < \alpha < 1,
\end{equation}
where \(\alpha\) is the forgetting factor, controlling the contribution of past interference values. A smaller \(\alpha\) assigns more weight to historical interference, making the estimator slower to react to rapid changes. Conversely, a larger \(\alpha\) allows the estimator to be more responsive to recent observations. The MA-based approach highlights its limitations in environments with non-stationary interference or abrupt fluctuations, where the GPR-based method excels in adaptability and uncertainty quantification.

 \begin{figure*}  
    \centering  
    \begin{subfigure}{0.329\linewidth}  
        \includegraphics[width=\linewidth]{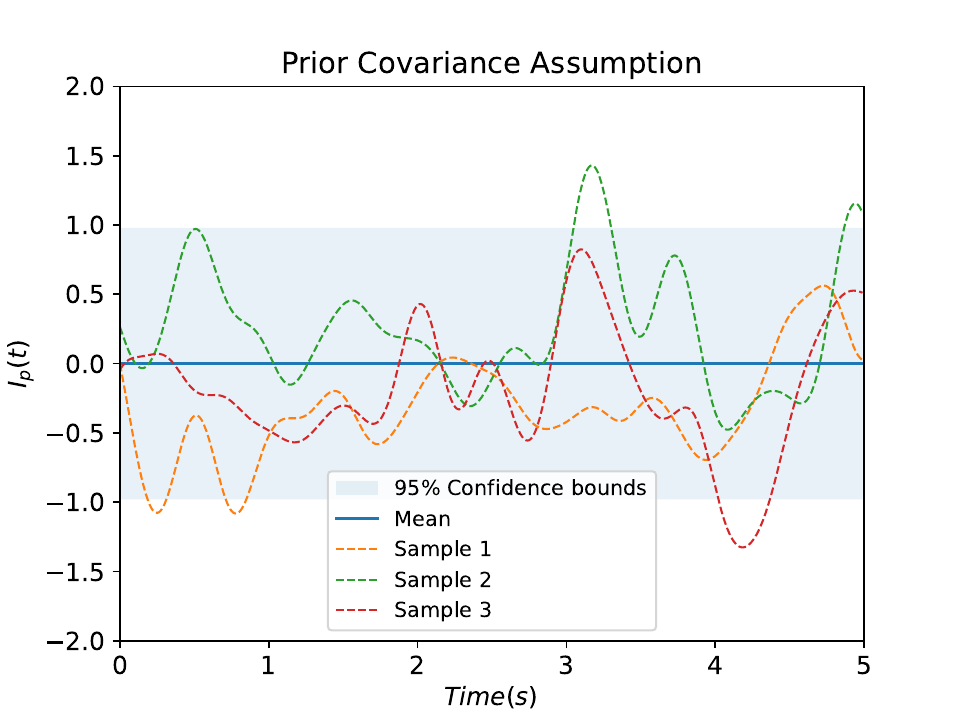}  
        \caption{}  
        \label{fig_p:enter-label1}  
    \end{subfigure}  
    \begin{subfigure}{0.329\linewidth}  
        \centering  
        \includegraphics[width=\linewidth]{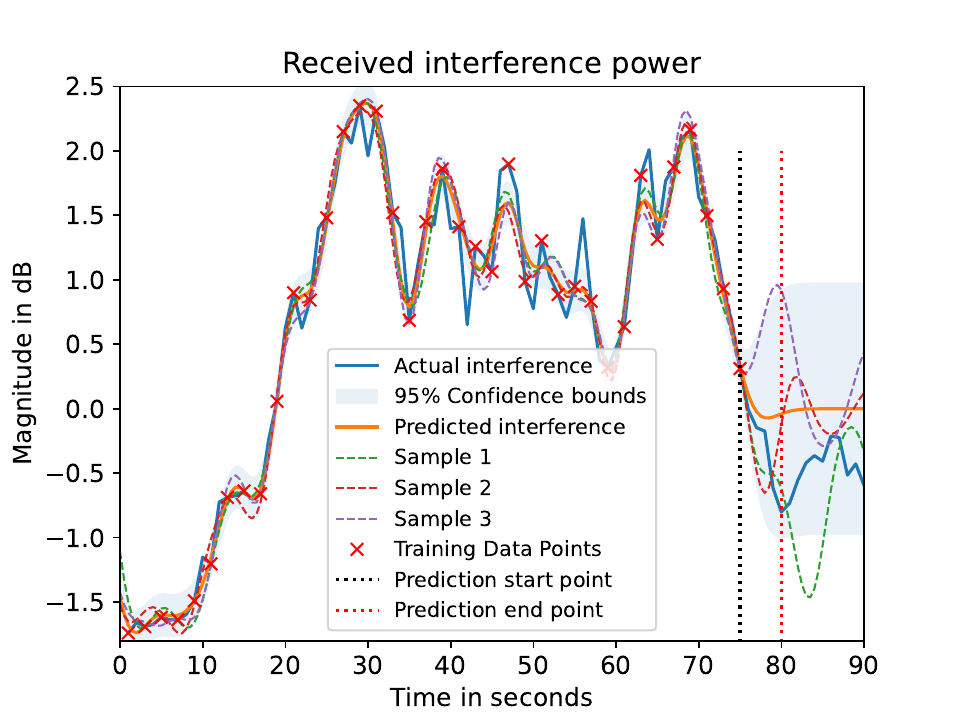}  
        \caption{}  
        \label{fig_p:enter-label2}  
    \end{subfigure}
    \begin{subfigure}{0.329\linewidth}  
        \includegraphics[width=\linewidth]{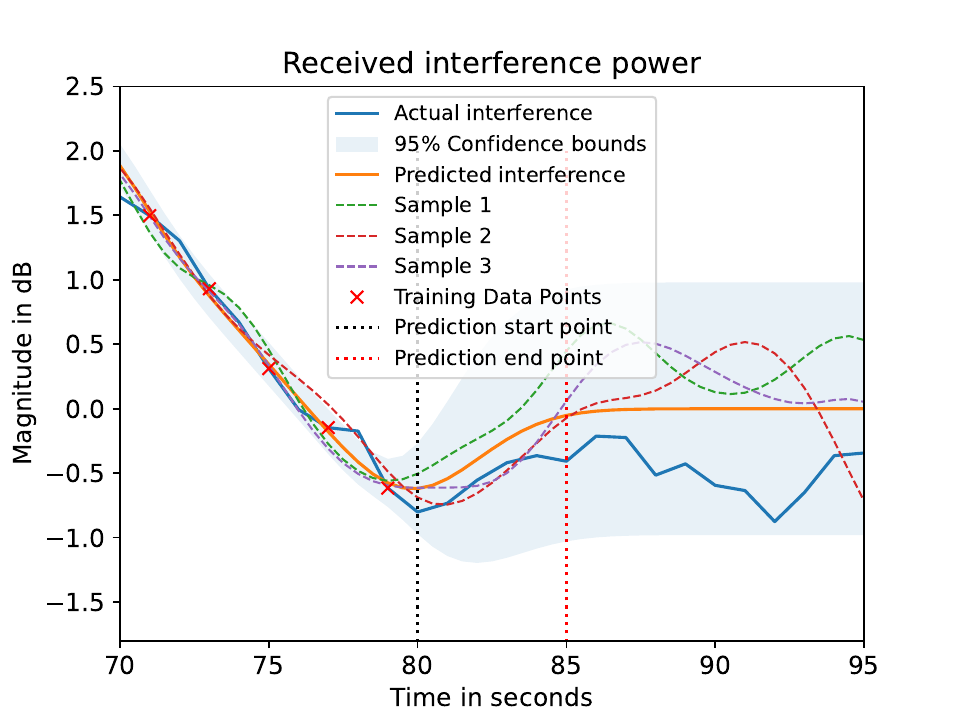}  
        \caption{}  
        \label{fig_p:enter-label3}  
    \end{subfigure}  
    \begin{subfigure}{0.329\linewidth}  
        \centering  
        \includegraphics[width=\linewidth]{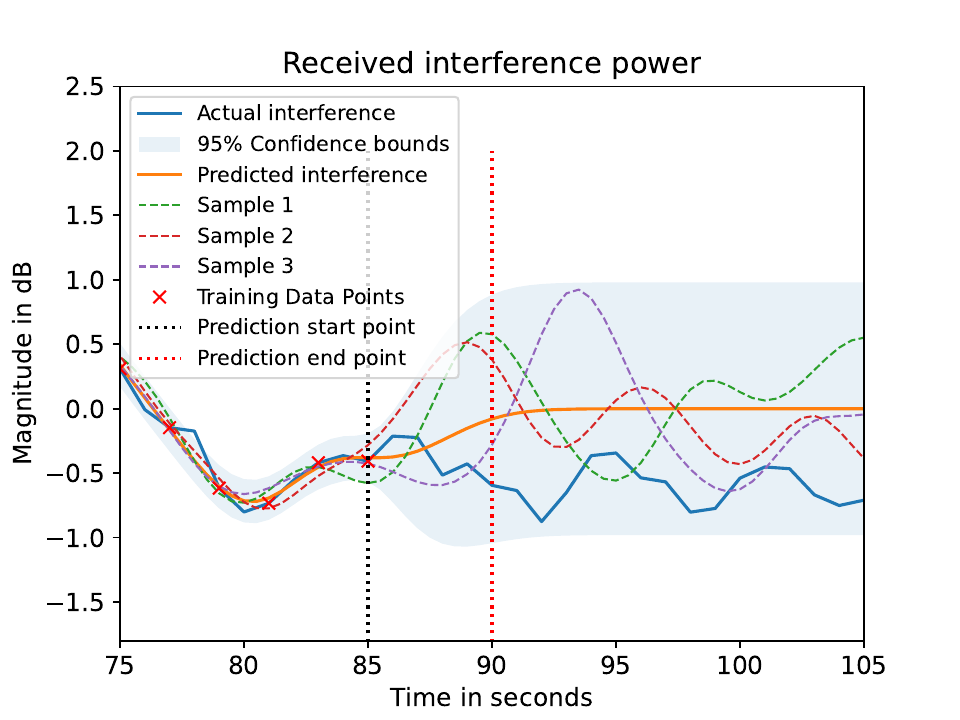}  
        \caption{}  
        \label{fig_p:enter-label4}  
    \end{subfigure}
    \begin{subfigure}{0.329\linewidth}  
        \includegraphics[width=\linewidth]{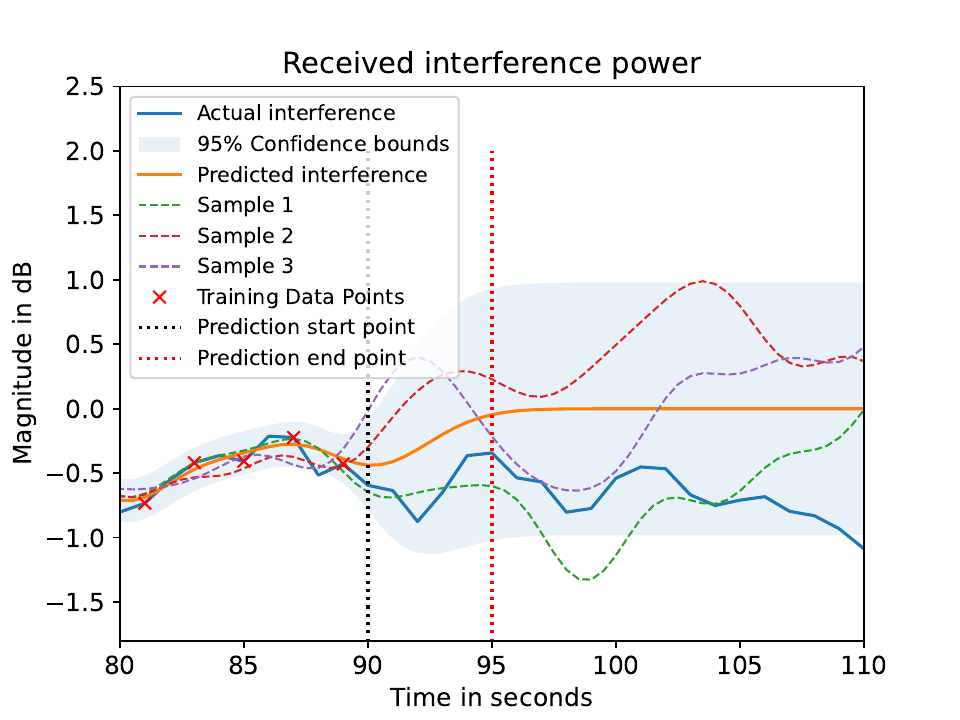}  
        \caption{}  
        \label{fig_p:enter-label5}  
    \end{subfigure}  
    \begin{subfigure}{0.329\linewidth}  
        \centering  
        \includegraphics[width=\linewidth]{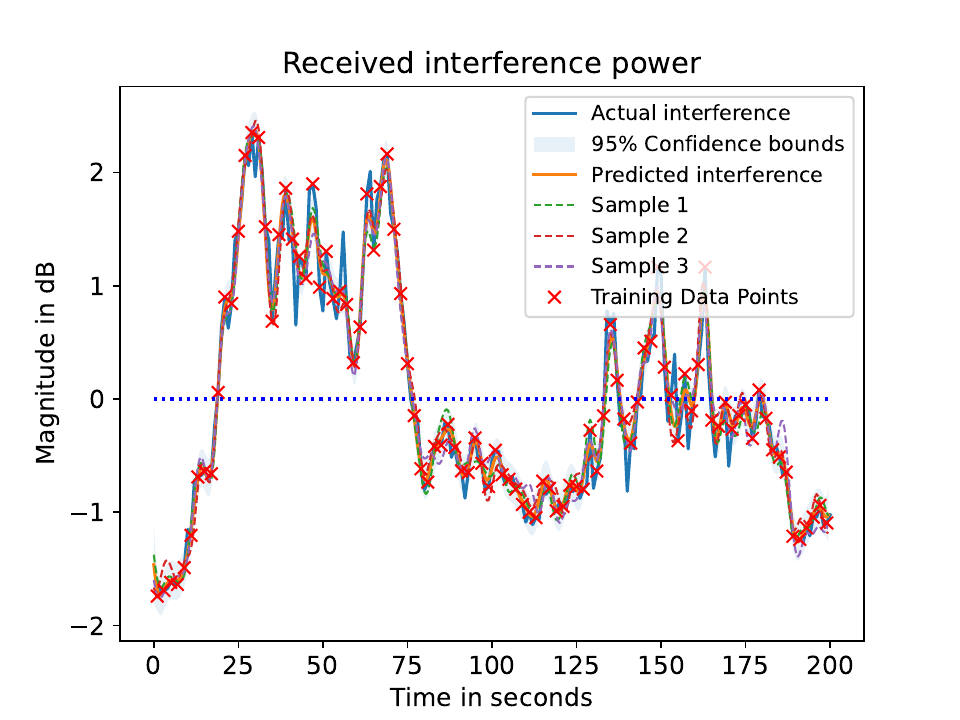}  
        \caption{}  
        \label{fig_p:enter-label6}  
    \end{subfigure} 
    \caption{Illustration of the GPR interference prediction method: (a) depicts prior beliefs before any data is observed, while (b)-(f) show the refinement of the model’s predictions and confidence bounds as actual interference data is incrementally observed and used to update future predictions.}
    \label{fig_p:overall-label} 
\end{figure*} 

\section{Results and Discussion}
\label{Sec: Results and Discussion}
This section discusses the obtained results. We generate the interference and desired signal based on the parameters summarized with other parameters in Tab.~\ref{tab1} for evaluating the GPR prediction performance~\cite{Samitha_Ref, Samitha_paper}.

\begin{table}[htbp]
\caption{Initial simulation parameters}
\begin{center}
\begin{tabular}{p{5cm} p{3cm}}  
\hline
\textbf{Parameter}  & \textbf{Value} \\
\hline
 SINR value of desired signal ($\bar{\gamma}$) & 20 dB \\
Number of interfered signals (N) & 6 \\
 Output scale ($\sigma_f$) & 0.5 \\
 INRs of interferers ($\gamma_i$ (dB)) & 5, 2, 0, -3, -10, 1 \\ 
 Channel model & Rayleigh block fading \\
 Length scale ($\ell$) & 2.5 \\
 Target error rates ($\varepsilon$) & [$10^{-5}, 10^{-4},$ $10^{-3}, 10^{-2}, 10^{-1}$] \\
 Gaussian noise ($\epsilon$) & $10^{-3}$ \\
  Forgetting factor ($\alpha$) & 0.01 \\
  Number of bits for resource allocation ($D$) & 50 \\
\hline
\end{tabular}
\label{tab1}
\end{center}
\end{table}

\subsection{GPR Interference Prediction and Uncertainty Analysis}
\label{subsec: GPR interference}
Based on the values of hyperparameters for the RBF kernel, outlined in Tab.~\ref{tab1}, the GPR model starts with a prior where the $I_p$ has a mean value of zero. In the initial phase, prior to the observation of any data, the model operates under the assumption of a Gaussian prior distribution with zero mean and 95\% confidence bounds, as depicted in Fig.~\ref{fig_p:enter-label1}. The confidence interval in this phase reflects the inherent uncertainty in interference prediction when no prior information is available. The three sample paths drawn from this prior distribution illustrate examples of possible interference power distributions, characterized by wide uncertainty due to the absence of any observed data.  

Once the training data points, denoted by the red crosses in Fig.~\ref{fig_p:overall-label}, are incorporated into the GPR model, a posterior update occurs. Specifically, the mean interference prediction is adjusted to match the actual observed interference samples using Eqs. (\ref{eq: 6}) and (\ref{Eqn: Posterior mean and variance}), as shown in Fig.~\ref{fig_p:enter-label2}. The update on the hyperparameters of the RBF kernel in this stage reduces the uncertainty, which is manifested as a narrowing of the $95\%$ confidence bounds. This reflects the model's improved understanding of the interference pattern based on the available data, as elaborated in Section~\ref{Sec: Regression Setting using GPR}.  

Beyond time step $75$, the model enters a predictive mode, where future interference values are predicted based on the updated posterior distribution. Fig.~\ref{fig_p:enter-label2} shows the $I_p$ for the interval from $75$ to $80$ seconds, which is derived from the posterior mean. Following the successful data transmission during this interval, the actual interference power at time step $80$ is observed and integrated into the model as additional training data. This triggers another posterior update, leading to refined predictions for the subsequent interval from $80$ to $85$ seconds, as demonstrated in Fig.~\ref{fig_p:enter-label3}.  

The sliding window process iteratively updates the GPR model as new interference measurements are received. At each time step, such as $85$ seconds, the model incorporates the latest observed interference, refining its posterior and enhancing prediction accuracy for subsequent intervals (e.g., $85$ to $90$ seconds), as illustrated in Fig.~\ref{fig_p:enter-label4}. This cycle continues, with each new observation incrementally improving the prediction accuracy, as shown for the $90$ to $95$ second interval in Fig.~\ref{fig_p:enter-label5}.  

The GPR-based sliding window approach enables continuous prediction of interference over a short time horizon, incorporating real-time interference measurements and dynamically updating the posterior distribution. This adaptive mechanism effectively balances the trade-off between prediction accuracy and uncertainty reduction, as highlighted in Fig.~\ref{fig_p:enter-label6}. The predicted interference values, together with the confidence bounds, provide valuable insights for proactive resource allocation in real-time, as discussed in the subsequent subsections. 

\subsection{Resource Allocation: GPR vs. Benchmark Schemes}
\begin{figure}
    \centering
    \includegraphics[width=1\linewidth]{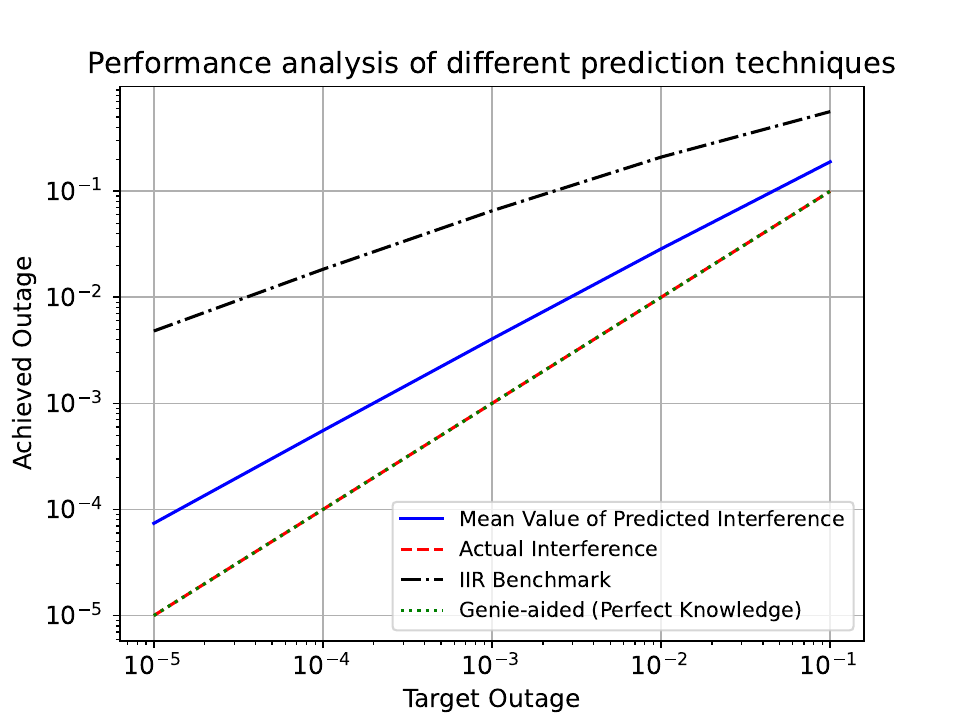}
    \caption{Achieved outage vs. target outage for different prediction techniques.}
    \label{Fig: ResourceAllocation_Results}
\end{figure}

In this section, we evaluate resource allocation for the downlink channel based on ${I_p}$. The key performance metrics used are the target outage probability ($\varsigma$), which represents the desired block error rate (BLER), and the achieved outage probability ($\hat{\varsigma}$), which measures the actual BLER after resource allocation. Two benchmark schemes are considered for comparison: the MA-based estimator and the genie-aided estimator. As shown in Fig.~\ref{Fig: ResourceAllocation_Results}, its resource allocation curve aligns closely with the actual interference, representing the best possible performance. Deviations from this curve indicate inefficiency in resource usage. The MA-based estimator, which uses an IIR filter for interference prediction, shows suboptimal performance. It achieves a BLER target of approximately $10\%$, but the achieved outage curve deviates significantly from the genie-aided baseline. While this method may be sufficient for eMBB services, it fails to meet the stricter outage requirements of HRLLC critical services. In contrast, the proposed GPR-based resource allocation scheme achieves near-optimal performance by adapting to real-time interference variations and incorporating uncertainty estimates. As depicted in Fig.~\ref{Fig: ResourceAllocation_Results}, the GPR model closely approaches the performance of the genie-aided estimator, allocating resources more efficiently than the MA-based estimator while maintaining low outage probabilities. This demonstrates the GPR model's ability to balance resource efficiency and reliability, especially in dynamic interference environments.

\subsection{Critical Discussion}  
The results demonstrate the clear advantage of the GPR-based approach for both interference prediction and resource allocation compared to the benchmark methods. Its sliding window framework ensures continuous adaptation to changing interference conditions, enabling real-time updates and dynamic resource reallocation. This adaptability is critical for HRLLC applications in 6G networks, where stringent reliability and low-latency requirements must be met.  

The GPR model provides confidence bounds for each prediction, as shown in Fig.~\ref{fig_p:overall-label}, with values consistently within the \( 95\% \) confidence interval. This is crucial for mission-critical services, offering quantifiable certainty regarding unpredictable processes like interference power variations, channel fluctuations, and packet arrivals at the MAC layer. Hyperparameter tuning further optimizes GPR performance, enhancing prediction accuracy. Though not perfect, the GPR-based method achieves near-optimal resource allocation in uncertain, variable interference conditions, making it a practical solution for HRLLC services.

\section{Conclusions and Potential Future Directions}  
\label{Sec: Conclusions and Future}  
This paper presents a GPR-based model for predictive interference management and resource allocation in 6G networks. The model effectively forecasts interference power for near-future transmissions, enabling proactive resource allocation that adapts to real-time channel conditions. By leveraging uncertainty estimates in interference predictions, the GPR model outperforms the traditional MA-based estimator and achieves near-optimal resource allocation performance, approaching the genie-aided benchmark. This demonstrates the model's ability to meet stringent QoS requirements, particularly for mission-critical HRLLC applications. Future work could focus on extending the GPR model to account for multi-user interference and spatial-temporal correlations in dense network environments. Furthermore, optimizing GPR hyperparameters for self-optimizing resource allocation could further enhance the system’s adaptability to dynamic interference conditions.
\section*{Acknowledgment}
The research leading to this paper was supported by the Research Council of Finland (former Academy of Finland) through the projects \href{https://www.6gflagship.com/}{6G Flagship} (grant number: 369116), and 6G-ConCoRSe (grant number: 359850).



\end{document}